\documentclass{rspublic}

\usepackage{graphicx}
\usepackage{amsmath,amssymb,amsfonts}
\usepackage{epstopdf}

\newcommand\al{\alpha}
\newcommand\be{\beta}
\newcommand\ga{\gamma}
\newcommand\de{\delta}
\newcommand\ep{\epsilon}

\newcommand\et{\eta}

\newcommand\la{\lambda}
\newcommand\rh{\rho}
\newcommand\si{\sigma}
\newcommand\ta{\tau}

\newcommand\ph{\phi}
\newcommand\vp{\varphi}
\newcommand\ch{\chi}

\newcommand\Ga{\Gamma}

\newcommand\pa{\partial}

\newcommand\ie{\emph{i.e.}}
\newcommand\eg{\emph{e.g.}}
\newcommand\ea{\emph{et al}}
\newcommand{\klmt}{\mbox{K\hspace{-6.5pt}KLM\hspace{-8.25pt}MT}\ }

\newcommand\beq{\begin{equation}}
\newcommand\eeq{\end{equation}}
\newcommand\bea{\begin{eqnarray}}
\newcommand\eea{\end{eqnarray}}

\newcommand\rms[1]{_{\text{#1}}}

\newcommand\half{\tfrac{1}{2}}
\newcommand\ap{\approx}
\newcommand\cd{\cdot}

\newcommand\X{\times}
\newcommand\fr{\frac}

\newcommand\Z{{\mathbb{Z}}}

\newcommand\gap{\;\lower2pt\hbox{$\buildrel > \over
{\scriptstyle\sim}$}\;}
\newcommand\lap{\;\lower2pt\hbox{$\buildrel < \over 
{\scriptstyle\sim}$}\;}
\newcommand\<{\langle}
\renewcommand\>{\rangle}

\newcommand\cM{\mathcal{M}}

\newcommand\bone{\mathbf{1}}

\newcommand\ba{\mathbf{a}}
\newcommand\bb{\mathbf{b}}
\newcommand\bp{\mathbf{p}}
\renewcommand\bq{\mathbf{q}}

\newcommand\bx{\mathbf{x}}
\newcommand{\OL}[1]{ \hspace{1pt}\overline{\hspace{-.5pt}#1
    \hspace{-1.5pt}}\hspace{1.5pt} }

\begin{document}

\title[Cosmic Strings]{
Cosmic Strings and Superstrings
}

\author[E.~J.~Copeland \& T.~W.~B.~Kibble]{Edmund J.~Copeland$^1$ \& T.~W.~B.~Kibble$^2$}
\affiliation{
$^1$School of
Physics and Astronomy, University of Nottingham,\\ University Park,
Nottingham NG7 2RD, United Kingdom\\
$^2$Blackett Laboratory, Imperial College, London SW7 2AZ, United 
Kingdom
\\ \rm Preprint number: Imperial/TP/09/TK/03
}

\maketitle

\begin{abstract}{cosmic strings, superstrings, early universe, branes, braneworlds}
Cosmic strings are predicted by many field-theory models, and may have been formed at a symmetry-breaking transition early in the history of the universe, such as that associated with grand unification.  They could have important cosmological effects.  Scenarios suggested by fundamental string theory or M-theory, in particular the popular idea of brane inflation, also strongly suggest the appearance of similar structures.  Here we review the reasons for postulating the existence of cosmic strings or superstrings, the various possible ways in which they might be detected observationally, and the special features that might discriminate between ordinary cosmic strings and superstrings.
\end{abstract}

\section{Introduction}
\label{intro}

There are many examples of linear topological defects in low-temperature con\-densed-matter systems, including vortices in superfluid helium and in atomic Bose-Einstein condensates, and magnetic flux tubes in superconductors.  These structures typically appear when a system goes through a phase transition into a low-temperature ordered phase in which the underlying symmetry is spontaneously broken.  They arise because in certain circumstances the ordering may be frustrated.

Cosmic strings are analogous objects that may have been formed in the early universe.  They could be produced during one of the early symmetry-breaking phase transitions predicted by many particle-physics models, for example one associated with the breaking of a `grand unification' symmetry.  Similar point-like or planar defects --- monopoles or domain walls --- may also be formed.

Superstrings are the supposed basic constituents of matter in fundamental string theory or M-theory, which is at present the leading contender as a unified theory of all particle interactions including gravity.

Both cosmic strings and superstrings are still purely hypothetical objects.  There is no direct empirical evidence for their existence, though there have been some intriguing observations that were initially thought to provide such evidence, but are now generally believed to have been false alarms.  Nevertheless, there are good theoretical reasons for believing that these exotic objects do exist, and reasonable prospects of detecting their existence within the next few years.

Initially, cosmic strings and superstrings were regarded as two completely separate classes.  Cosmic strings, if they exist, stretch across cosmological distances and though exceedingly thin are sufficiently massive to have noticeable gravitational effects.  On the other hand superstrings were seen as minute objects, even on the scale of particle physics, far too small to have any directly observable effects.  But in the last few years it has emerged that under certain circumstances they too can grow to macroscopic size and play the same role as cosmic strings.  Indeed, it is now very possible that observations of a cosmic superstring might provide the first real direct evidence for M-theory.  (For reviews see  Polchinski 2005, Davis \& Kibble 2005, Sakellariadou 2009, Myers \& Wyman 2009.)

We begin, in Section \ref{cosmicstr}, by describing the simplest examples of cosmic strings, which arise in the breaking of an abelian U(1) symmetry, and discuss the reasons for thinking that cosmic strings might appear during phase transitions in the very early history of the universe, in a process closely analogous to what happens when a superfluid or superconductor is cooled rapidly through its transition temperature.  In Section \ref{dynamics} we review the dynamics of cosmic strings, and in Section \ref{evolution}, discuss how a network of cosmic strings, once formed, would evolve as the universe expands.  

In Section \ref{sustr}, we review the ideas behind fundamental superstring theory and the characteristics of the five known consistent superstring theories.  In all five cases, consistency requires a spacetime of ten rather than four dimensions.  A mechanism is therefore required to reduce the number of dimensions of our spacetime to the familiar four.  One such is the Kaluza--Klein mechanism of \emph{compactification}: the unobserved dimensions are in fact curled up so small as to be unobservable on ordinary scales.  Another idea, the \emph{brane-world} mechanism, has emerged more recently.  A key feature of superstring theories is the appearance of \emph{branes}, surfaces of various dimensions embedded in the larger spacetime.  A $p$-brane is a surface of spatial dimension $p$, or spacetime dimension $p+1$.  The basic idea of the brane world is that all the constituents of our universe are bound to a 3-brane that can move within the larger nine-dimensional space; they in fact consist of short segments of fundamental string with their ends firmly rooted on the brane.

In Sections \ref{tension}-\ref{stability}, we explain why the initial rejection of the idea of cosmic superstrings has been overturned by showing how they can have sufficiently small tensions, survive a period of inflation and be cosmologically stable, and use a series of models to demonstrate the point, using them to also demonstrate the similarities and differences  between these cosmic superstrings and ordinary cosmic strings (often called \emph{solitonic} strings).  One particularly important difference, discussed in Section \ref{reconnect}, lies in the probability of `intercommuting' or exchange of partners when strings cross, and also in the possibility of forming three-string junctions. The observational signatures of the different types of strings are the subject of Section \ref{observation}, where we also discuss the current observational bounds on their existence.

Our conclusions are summarized in Section \ref{conc}. 

\section{Cosmic strings}
\label{cosmicstr}

A key feature of particle physics over the last few decades has been the search for \emph{unification}.  A cornerstone of the currently accepted Standard Model is the \emph{electroweak} theory which provides a unified description of the weak and electromagnetic interactions, and for which Glashow, Salam and Weinberg shared the 1979 Nobel Prize for Physics.  It is a gauge theory based on the symmetry group SU(2)$\X$U(1).  At high  energies, this symmetry is manifest, but at energies below a few hundred GeV, it is hidden, broken spontaneously down to the residual U(1) subgroup representing electromagnetic gauge transformations.  Of the four gauge bosons in the theory, the three that mediate weak interactions, the $W^\pm$ and $Z^0$, acquire a mass of order 100 GeV, while the fourth, the photon, remains massless.

There is good reason to believe that some kind of unification occurs at a much higher energy scale between the strong interactions and the electroweak ones.  The coupling `constants' do in fact vary slowly with energy, and all of them approximately converge to a common value at high energy.  The convergence is much more precise in the minimal supersymmetric extension of the Standard Model, provided that supersymmetry breaking occurs at a fairly low energy, around 1 TeV.  Then all the couplings coverge at a \emph{grand unification} scale somewhere in the region of $10^{16}$ GeV (Amaldi \ea\ 1992).  Several \emph{grand unified theories} (\emph{GUTs}) have been proposed, in which some larger symmetry group $G$ would be effective at extremely high energies, but is spontaneously broken at the GUT scale down to the symmetry group of the Standard Model, namely SU(3)$\X$SU(2)$\X$U(1), where the SU(3) is the symmetry group of quantum chromodynamics, which describes the strong interactions, and SU(2)$\X$U(1) is the electroweak group.

\begin{figure}
\begin{center}
  \includegraphics[width=2in,angle=0]{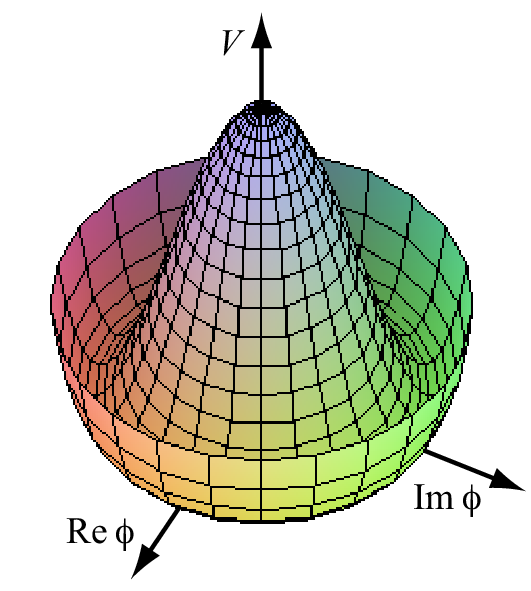}
\caption{The sombrero potential.}
\label{sombrero}
\end{center}
\end{figure}
Spontaneously broken gauge theories such as these often exhibit stable \emph{topological defects} --- domain walls, cosmic strings or monopoles.  (See Jeanerrot \ea\ 2003.)  The simplest model in which cosmic strings arise is a U(1) gauge theory, involving a complex scalar field interacting with the electromagnetic field and described by the Lagrangian density
 \beq 
 L=D_\mu\ph^*D^\mu\ph-\tfrac{1}{4}F_{\mu\nu}F^{\mu\nu}
 -V(\ph). 
 \eeq
Here the covariant derivative and electromagnetic field are $D_\mu\ph=\pa_\mu\ph+ieA_\mu\ph$ and $F_{\mu\nu}=\pa_\mu A_\nu-\pa_\nu A_\mu$, $V(\ph)$ is the \emph{sombrero potential} $V=\half\la(\ph^*\ph-\half\et^2)^2$, in which $\la$ and $\et$ are positive constants (see Fig.~\ref{sombrero}), and we use natural units with $c=\hbar=k\rms{B}=1$.  This model is invariant under the local U(1) gauge transformations
 \bea 
 \ph(x)&\to&\ph(x)e^{i\al(x)},\notag\\
 A_\mu(x)&\to&A_\mu(x)-\fr{1}{e}\pa_\mu\al(x). 
 \label{gaugetrans} 
 \eea

At high temperatures, there will be large fluctuations of $\ph$ around a zero average.  However, when the temperature falls below a critical value $T\rms{c}$ of order $\et$, $\ph$ will tend to settle down into the valley of the potential.  At zero temperature, we expect $\<\ph\>=\et e^{i\al}$ with arbitrary phase angle $\al$.  The theory thus exhibits \emph{spontaneous symmetry breaking}: when the temperature falls below $T\rms{c}$, the field $\ph$ must randomly choose a phase angle.  Moreover this choice may be made differently in widely separated regions, so the field may be frustrated from reaching the minimum everywhere.  In particular, if around a large loop $\al$ varies by $2\pi$ then somewhere within that loop, $\ph$ must vanish; the locus of such points is a cosmic string.

A straight cosmic string along the $z$-axis is a solution of the form (Nielsen \& Olesen 1973)
 \bea 
 \ph(t,z,\rh,\vp)&=&\fr{\et}{\surd 2}f(\rh) e^{in\vp},
 \notag\\
 A_\mu(t,z,\rh,\vp)&=&-\fr{n}{e}h(\rh)\pa_\mu\vp, 
 \eea
where we use cylindrical polar coordinates.  The functions $f$ and $h$ are determined by minimizing the energy subject to the boundary conditions
 \bea f(0)=0,\ f(\infty)=1,\qquad h(0)=0,\ h(\infty)=1.\eea
(See for example Hindmarsh \& Kibble 1994, Vilenkin  \& Shellard 1994).
 
The solution with $n=\pm 1$ is stable for topological reasons, because around a circle at infinity the phase of $\ph$ changes by $2\pi$, so $\ph$ follows an incontractible path on the manifold of minima of $V$, namely the circle $|\ph|=\et/\sqrt 2$.  For these relativistic strings, the energy per unit length is equal to the string tension $\mu$, which is related to the scale $\et$ by 
 \beq \mu=F(\be)\et^2, \label{mu} \eeq
where $F$ is a function of order 1 of the ratio $\be$ between the gauge and scalar couplings, which also turns out to be the ratio between the squared masses of the vector and scalar particles:
 \beq \be=\fr{e^2}{\la}=\fr{m^2\rms{v}}{m^2\rms{s}}.
 \label{beta} \eeq

More generally, consider a theory with gauge symmetry group $G$ involving a multiplet $\ph$ of scalar fields, where the Lagrangian is invariant under transformations of the form $\ph(x)\to U(x)\ph(x)$ for any $U\in G$ (together with appropriate transformations of the gauge fields).  In the vacuum state, the expectation value $\<0|\ph(x)|0\>=\ph_0$, say, must lie on the manifold $\cM$ on which the potential $V(\ph)$ takes its minimum value (analogous to the circle $|\ph|=\et/\sqrt 2$ above).  In general, the symmetry is spontaneously broken, but only partially.  The unbroken symmetry subgroup $H$ is the stability group of $\ph_0$, namely
 \beq H=\{U\in G: U\ph_0=\ph_0\}. \eeq
The degenerate minima, the points of $\cM$, are labelled by the left cosets of $H$ in $G$.  Equivalently, $\cM$ may be identified with the quotient space
 \beq  \cM=G/H, \eeq
whose elements are precisely these cosets.

It is the topology of this manifold $\cM$ that determines whether or not topological defects such as cosmic strings can appear (see for example Kibble 2000).  When the system cools below the relevant symmetry-breaking scale the value of $\ph$ will tend to settle down everywhere towards some point on the manifold $\cM$, but the choice of point may vary from place to place, and will be made independently in widely separated regions, so the symmetry breaking may be frustrated.  In particular, cosmic strings are found if and only there are incontractible loops within $\cM$, \ie\ if $\cM$ is not simply connected, or equivalently if its fundamental group or first homotopy group is non-trivial: $\pi_1(\cM)\ne\bone$ ($\bone$ is the group comprising the identity only).  In that case, if the values of $\ph$ around some loop in space follow an incontractible loop in $\cM$ then in the interior of the loop $\ph$ must leave the manifold $\cM$; a cosmic string, or strings, must pass through the loop.

In such a case, the topologically inequivalent classes of cosmic string are classified by the elements of the fundamental group.  For the simple case of the breaking of U(1) down to $\bone$, $\cM$ is a circle $S^1$, and the fundamental group is the group of integers, $\pi_1(\cM)=\Z$.  Here the possible strings are labelled by an integer, the \emph{winding number}; for a typical string of winding number $n$, the phase of $\ph$ around the string is $\al=n\vp$.  What this analysis does not tell us is whether all these strings are stable.  The $n=1$ string is always stable, but for higher values of $n$ it may be energetically favourable for a string with winding number $n$ to break up into $n$ separate strings of winding number 1.  As a matter of fact, whether this is so or not depends on the parameter $\be$ defined in (\ref{beta}).  If $\be>1$, strings with all values of $n$ are stable; if $\be<1$, only the $n=\pm 1$ strings are stable.  This is the same distinction as that between Type-I and Type-II superconductors.

It is also possible to have strings in theories with no gauge field, where the symmetry is global rather than a local, gauge symmetry as in (\ref{gaugetrans}).  These are called \emph{global}, as opposed to \emph{local}, strings, and have rather different properties.  We shall not consider them further here.

In many models of unification the symmetry group $G$ is simple or semi-simple.  In such cases, the fundamental group $\pi_1(\cM)$ is always a finite group.  For example, suppose that $\ph$ is a traceless symmetric tensor under SO(3), belonging to the five-dimensional $j=2$ representation, and that $\ph_0=(\et/\sqrt{6})\mathrm{diag}(-1,-1,2)$.   Then it is easy to see that $H=$~O(2), and one finds that $\pi_1(\cM)=\Z_2$, the group of integers modulo 2.  In this case, there is only one non-trivial class of cosmic string.

Very similar criteria determine whether or not other types of topological defects can exist.  Domain walls exist if the manifold $\cM$ is not connected, \ie\ if $\pi_0(\cM)\ne\bone$, while monopoles exist if there are incontractible 2-surfaces, \ie\ if $\pi_2(\cM)\ne\bone$.

\section{Dynamics of strings}
\label{dynamics}

Let us suppose there is symmetry breaking at some large energy scale $\et$, and that the topological condition for the formation of strings, $\pi_1(\cM)\ne\bone$, is satisfied.  Then when the universe cools through this temperature, the result will be a random tangle of cosmic strings, characterized by some characteristic distance $\xi$ between them.  (Some authors use the symbol $L$ for $\xi$.)

Relativistic strings of the simplest type are characterized by the fact that in the string rest-frame, the energy per unit length is equal to the string tension $\mu$, given by (\ref{mu}).  We shall always use `length' of string to mean the invariant length, the energy divided by $\mu$.  The characteristic scale $\xi$ may be defined by saying that in a randomly chosen large cube of size $\xi^3$, the average total length of string is $\xi$, or equivalently the mean energy density in strings is
 \beq \rh\rms{str}=\fr{\mu}{\xi^2}. \label{rhostring}\eeq

For reasons of causality, the choice among the degenerate minima will be made independently in widely separated regions; the choices cannot be correlated over distances larger than the causal horizon distance (Kibble 1976), which in the early universe is comparable with its age $t$.  (For example, in a radiation-dominated universe, it is $2t$.)  This puts a lower limit on the initial density of cosmic strings: $\xi\lesssim t$, and in fact we might expect that $\xi$ would substantially smaller, because the relevant causal processes do not date back as far as the Big Bang; this is confirmed by simulations of string formation.

Because cosmic strings are so thin (with widths of order $1/m\rms{s}=1/(\sqrt{\la}\et)$ or $1/m\rms{v}=1/(e\et)$) their large-scale dynamics in an otherwise empty or low-density universe is very well described by the \emph{Nambu-Goto action}, a zero-width approximation.  The spacetime coordinates of points on the string are $x^\mu(\si^a)$, where $\si^a$ ($a=0,1$) are two world-sheet coordinates, and the action is required to be invariant under arbitrary reparametrizations $\si^a\to\si^a{}'$.  In fact the Nambu-Goto action is simply proportional to the area of the world-sheet:
 \beq S=-\mu\int d^2\si \sqrt{-\det(g_{\mu\nu}x^\mu_{,a}x^\nu_{,b})}, \eeq
in which $g_{\mu\nu}$ is the spacetime metric and $x^\mu_{,a}=\pa x^\mu/\pa\si^a$.

It is often convenient to choose one of the world-sheet coordinates to be the global time coordinate: $\ta=\si^0=x^0$.  In Minkowski space, the other coordinate $\si=\si^1$ may be chosen to satisfy also the `conformal gauge' conditions
 \beq \dot x^2+(x')^2=0, \qquad \dot x\cd x'=0, \eeq 
where $\dot x=\pa x/\pa\ta$ and $x'=\pa x/\pa\si$, or equivalently in terms of three-vectors
 \beq \dot \bx^2+(\bx')^2=1, \qquad \dot\bx\cd\bx'=0.
 \label{gaugecon} \eeq 
Then it turns out that the Nambu-Goto equation of motion is just the two-dimen\-sion\-al wave equation,
 \beq \ddot\bx=\bx''. \eeq
So in this case the general solution is easy to write down; it is a linear superposition of left-moving and right-moving waves, each moving along the string with the speed of light:
 \beq \bx(\ta,\si)=\half[\ba(\si+\ta)+\bb(\si-\ta)], \eeq
where to satisfy (\ref{gaugecon}) both $\ba'$ and $\bb'$ must be unit vectors, 
 \beq \ba'{}^2=\bb'{}^2=1.\label{a'b'} \eeq 

In an expanding (spatially flat) Friedmann-Lema\^\i tre-Robertson-Walker spacetime, with metric
 \beq ds^2 = a^2(\ta)(d\ta^2 - d\bx^2), \eeq
these conditions are no longer compatible.  It is still possible to impose the `transverse gauge' condition $\dot\bx\cd\bx'=0$, but then the quantity
 \beq \ep(\ta,\si)=\sqrt{\fr{\bx'{}^2}{1-\dot\bx^2}}, \eeq
is no longer a constant, and the equations take the more complicated form
 \bea \ddot\bx+2\fr{\dot a}{a}(1-\dot\bx^2)&=&
 \fr{1}{\ep}\left(\fr{1}{\ep}\bx'\right)',\label{NGeqn} \\
 \dot\ep&=&-2\fr{\dot a}{a}\ep\dot\bx^2. 
 \notag \eea
These equations do not have non-trivial exact solutions but are still numerically tractable.  Here too the motion can be thought of as a superposition of left and right moving waves.  The analogues of $\ba'$ and $-\bb'$ are the unit vectors 
 \beq \bp=\dot\bx+\ep^{-1}\bx', 
 \qquad \bq=\dot\bx-\ep^{-1}\bx'.
 \label{pq} 
 \eeq
The difference is that now the left- and right-movers \emph{do} interact.  It is useful to introduce a null world-sheet coordinate $u$ constant along the left-moving null geodesics, \ie\ $\dot u=\ep^{-1}u'$, and to regard $\bp$ as a function of $\ta$ and $u$ (and similarly $\bq$ as a function of $\ta$ and $v$, where $\dot v=-\ep^{-1}v'$).   Then one finds that the dependence of $\bp,\bq$ on $\ta$ is slow: they satisfy the equations of motion
 \bea \pa_\ta\bp(\ta,u)&=&-\fr{\dot a}{a}[\bq-(\bp\cd\bq)\bp], \\
 \pa_\ta\bq(\ta,v)&=&-\fr{\dot a}{a}[\bp-(\bp\cd\bq)\bq] \eea
Note that $\bp$ is nearly constant along a left-moving null geodesic; the characteristic time for change with $\ta$ is the Hubble time.

\begin{figure}
\begin{center}
  \includegraphics[width=3in,angle=0]{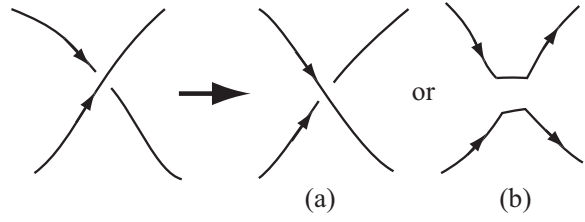}
\caption{Intercommuting of two strings.}
\label{intercommuting}
\end{center}
\end{figure}
There is however one question the Nambu-Goto equations cannot answer: what happens when two strings (or two segments of the same string) meet?  For the simplest types of string, two outcomes are possible (see Fig.~\ref{intercommuting}): (a) the strings might pass through one another, or (b) they might exchange partners, a process often called \emph{intercommuting}.  Field theory simulations of simple models have shown that cosmic strings almost always choose the second alternative (Shellard 1987, Matzner 1989); the probability of intercommuting is 1 except perhaps when the relative velocity of the two strings is very close to the speed of light (Hanany \& Hamimoto 2005, Ach\'ucarro \& De Putter 2006, Eto \ea\ 2006).

\begin{figure}
\begin{center}
  \includegraphics[width=3in,angle=0]{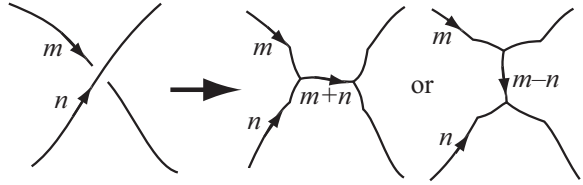}
\caption{Junctions formed by Type-I strings.}
\label{zip}
\end{center}
\end{figure}

However there are types of string for which other possibilities exist.  For example, for Type-I strings, where there are stable solutions with higher winding numbers, three-string junctions can form (Bettencourt \& Kibble 1994).  When strings of winding numbers $m$ and $n$ meet, they can zip together to create a third string of winding number $m\pm n$ (see Fig.~\ref{zip}).
There are in this case kinematic constraints on which of these possibilities is realized (see Copeland \ea\ 2007).  As is discussed later, some cosmic superstrings also exhibit this possibility.  Even when they are not junction-forming, the probability of intercommuting of superstrings is often much less than unity.

A particularly interesting situation occurs if $\pi_1(\cM)$ is a non-abelian group, such as the quaternion group, of order 8.  Here too there can be three-string junctions.  It turns out that strings labelled by non-commuting elements of  $\pi_1(\cM)$ may zip together but cannot simply pass through one another without becoming joined by a third string, labelled by the commutator.

\section{Evolution of a string network}
\label{evolution}

Now let us consider what happens to a tangle of cosmic strings once formed by early symmetry breaking.  Initially, at least in most scenarios, just after the time of formation, the universe will be dense, with an energy density only slightly less than that inside the strings, so their motion will be heavily damped.  If it were not for the universal expansion, for energetic reasons, the total length of string would tend to decrease as kinks become straightened under the effect of string tension.  From time to time, strings would meet and intercommute, forming new kinks that straighten out in turn.  On the other hand, the effect of expansion will be to cause the strings to lengthen.  It can be shown that during this initial \emph{friction-dominated} regime, the characteristic scale of the tangle will increase with time as $\xi\propto t^{5/4}$, so $\xi$ grows slightly faster than the Hubble radius (Kibble 1982).

Clearly this process cannot continue indefinitely.  For reasons of causality, $\xi$ can never exceed the causal horizon distance, of order $t$.  In fact, the friction ceases to be important when the typical damping time exceeds $\xi$, and this also occurs when $\ga=\xi/t$ has become roughly of order 1.  Thereafter, the strings move freely, and are well described by the Nambu-Goto action, reaching relativistic speeds --- in Minkowski space, the r.m.s.\ velocity is $1/\surd 2$, though in the expanding universe it is somewhat less.  There is now general agreement from several simulations, as well as theoretical analyses, that the string network will reach a \emph{scaling} regime in which the ratio $\ga=\xi/t$ remains nearly constant.  Estimates of this constant are typically of order 0.3 and 0.55 in the radiation- and matter-dominated eras respectively.  Equivalently, using (\ref{rhostring}), and remembering that the total energy density of the universe $\rh$ also behaves as $\rh\propto 1/t^2$, we see that in the scaling regime the energy density in strings remains a fixed fraction of the total energy density: if $a(t)\propto t^\nu$ ($\nu=\fr{1}{2}$ or $\fr{2}{3}$ in the radiation or matter era) then
 \beq 
 \fr{\rh\rms{str}}{\rh}=
 \fr{8\pi G\mu}{3\nu^2}\fr{\xi^2}{t^2}=
 \fr{8\pi G\mu\ga^2}{3\nu^2}, 
 \eeq
a constant of order $G\mu$, where $G$ is Newton's constant.  This dimensionless number $G\mu$ is of paramount importance in determining the observational effects of cosmic strings.

Any comoving volume in the universe is growing as $V\propto a^3 \propto t^{3\nu}$.  If the strings were fixed in comoving coordinates and just being stretched by expansion, we would have $\xi\propto a$ or $\rh\rms{str}\propto t^{-2\nu}$; strings would then eventually dominate the energy density of the universe.  Clearly, therefore, maintenance of the scaling regime requires some mechanism for removing energy from the tangle of strings.  It is widely believed that the formation of closed loops plays a key role, but the questions of how large those loops typically are and of the role of small-scale structure on the strings are still controversial.  This is a reflection of the fact that these questions involve such wide ranges of length and time scales that they are hard to deal with numerically.

When a string crosses itself, intercommutation can lead to the formation of a closed loop.  Isolated loops oscillate quasi-periodically, with a period equal to half the length $l$ of the loop.  If the loop is large ($l\gtrsim \xi$) then it is very likely to meet another piece of string and reconnect to the long-string network.  It may also intersect itself and break up into smaller loops, but if it survives then after a few oscillation periods we will be left with one or more long-lived loops, which oscillate and lose energy very gradually, by gravitational or other radiation.  It was initially assumed that the typical size of the loops lost to the network would be comparable to $\xi$.  But most of the early simulations of evolution using the Nambu-Goto equations found only very small loops with a typical size $l\ll \xi$ (Bennett \& Bouchet 1990, Allen \& Shellard 1990).

In estimating the likely observational effects of cosmic strings, a very important parameter is the typical size $l$ of the loops. This view that $l/t$ becomes very small has been supported by some recent high-resolution simulations (Martins \& Shellard 2006, Ringeval \ea\ 2007).  They suggest that the typical value of $l$ may actually be nearly constant, determined primarily by the initial conditions.  Other recent work, however, has suggested that there are two distinct families of loops, and that in addition to the very tiny loops a population of large loops will eventually appear (Vanchurin \ea\ 2006, Polchinski \& Rocha 2007, Vanchurin 2008, Dubath \ea\ 2008), typically with $l/t\sim 0.1$.  None of the simulations as yet covers a long enough period of time to be absolutely sure whether this is true.  It is very important to resolve this issue.

It has often been suggested that the small-scale structure on cosmic strings and the typical size of loop are determined by the effects of gravitational radiation.  However, the work of Siemens \& Olum (2001) has shown that gravitational radiation is much less efficient than had been thought, essentially because waves travelling in one direction do not interact strongly with all those in the opposite direction, but only those with a similar wavelength.  This implies that the loops will be smaller than earlier discussions had suggested.

The lifetime $t_l$ of a loop decaying by gravitational radiation will be roughly proportional to its size $l$.  The original estimate (Vachaspati \& Vilenkin 1985) was that
 \beq t_l \ap \fr{l}{\Ga G\mu}, \eeq
where $\Ga$ is a numerical factor depending on the shape, but not the size, of the loop, typically of order $10^2$.  But in some case, the lifetime may be substantially larger than this estimate.

There is also debate about whether gravitational radiation is indeed the primary energy-loss mechanism.  It has been suggested (Vincent \ea\ 1998, Hindmarsh \ea\ 2009) that the strings lose energy mainly by radiating other kinds of particles, associated with the gauge and scalar fields.  This too is an issue that needs to be resolved.

One reason for the production of small loops is that there is a great deal of small-scale structure on the strings.  Every time two strings meet and exchange partners, a pair of kinks is created on each of the strings, that propagate away from each other at the speed of light.  These kinks persist during the later evolution. The kink angle decreases, very slowly, due to the effect of stretching.  It is difficult to obtain information about this from the simulations, because the scales in question are close to the resolution limit.  Early analytic studies (Kibble \& Copeland 1991, Austin \ea\ 1995, Martins \& Shellard 1996) gave a good account of large-scale structure but were less successful on small scales.  However, considerable progress has been made recently in understanding the nature of the small-scale structure.  It is useful to discuss the correlation function 
 \beq 
 z(s,t) \equiv \<\bp(\ta,u_1)\cd\bp(\ta,u_2)\>, 
 \qquad s=a\ep(u_1-u_2), 
 \eeq
where the angle-brackets denote an ensemble average, and $t$ is the physical time defined by $dt = a(\ta)d\ta$.  On most scales of observational significance, the small-scale structure does reach a scaling regime, in which
 \beq
 z(s,t) = 1-A(s/t)^{2\ch},
 \eeq
where $A$ and $\ch$ are constants (Polchinski \& Rocha 2006).  On \emph{very} small scales, where $s$ is so small that at most one kink is likely to be found on the string segment, there is a non-scaling regime (Copeland \& Kibble 2009).  However as the number of kinks grows, the range of values of $s/t$ where this non-scaling applies shrinks with time.

\section{Cosmic superstrings}
\label{sustr}
String theory is based on the idea that the fundamental constituents of matter are not point particles but tiny strings, either open or closed (forming loops).  (For an introduction, see Polchinski 1998, Becker \ea\ 2007).  The theory can only be made consistent in a spacetime of more than the familiar four dimensions --- 26 for bosonic strings or 10 for superstrings which incorporate \emph{supersymmetry}, connecting bosons and fermions.  There are five known consistent superstring theories.  Type I is a theory of unoriented strings, while those of Type II are oriented; there are two distinct Type II theories, IIA and IIB, with differing symmetry properties.  Finally we have heterotic strings which are also oriented, and involve a curious combination of a left-moving bosonic string and a right-moving superstring; there are two such theories, based on the symmetry groups E$_8 \X $E$_8$ and SO(32) respectively.

One way of reducing the number of dimensions to the four we observe is the \emph{Kaluza-Klein} mechanism, in which it is assumed that the remaining six dimensions are \emph{compactified}, curled up to a small size, for example forming a loop of small radius $R$, just as a drinking straw appears one-dimensional when viewed on a large scale, though it is fundamentally two-dimensional.

Many remarkable duality relations have been established between these different superstring theories, for example a $T$-duality which relates a IIA theory compactified on a scale $R$ and a IIB theory compactified on a scale $l\rms{s}^2/R$, where $l\rms{s}$ is the fundamental string scale, related to the Planck length, as well as $S$-duality between theories with string coupling constant $g\rms{s}$ and $1/g\rms{s}$.  As a result, it has been established that all five superstring theories are limiting cases, in different regions of parameter space, of a single underlying 11-dimensional theory, called \emph{M-theory}.  So too is the maximal 11-dimensional supergravity theory, the supersymmetric extension of general relativity in 11 dimensions.

More recently, a different mechanism for recovering the observed four dimensions has emerged, the \emph{brane-world} mechanism, which is set in the context of Type-IIB string theory.   \emph{Branes} are surfaces of various dimensions embedded in the larger spacetime.  A $p$-brane is a surface of spatial dimension $p$, or spacetime dimension $p+1$.  Thus a 1-brane is a string, which sweeps out a two-dimensional \emph{world sheet} in spacetime; a 2-brane is a membrane (hence the name), sweeping out a three-dimensional \emph{world volume}; and so on.  In particular, there are  \emph{Dirichlet branes} (or \emph{D-branes}); fundamental strings (\emph{F-strings}) may either form closed loops or be open-ended;  a D$p$-brane is a ($(p+1)$-dimensional) surface on which the fundamental strings can terminate.  A D1-brane is called a D-string.  One important distinction between the two Type-II string theories is that stable D$p$-branes exist with even $p$ in Type IIA, and for odd $p$ in Type IIB.  D$p$-branes are generally coupled to $(p+1)$-form fields and carry corresponding fluxes.
 
The essence of the brane-world mechanism is that all the constituents of our universe are bound to a 3-brane that can move within the larger nine-dimensional space; they in fact consist of short segments of fundamental string with their ends firmly rooted on the brane.  This applies to all the particles of the Standard Model, including photons, so we cannot see beyond our 3-brane, but it does not apply to gravitons, which correspond to closed string loops.  This means that gravity can leak away from our brane, giving rise to possible observable deviations from Newtonian gravity on large scales.

In addition to F-strings and D-branes, other similar objects exist in M-theory, Neveu-Schwarz- (NS-)branes and M-branes.

The ideas of cosmic strings and superstrings emerged at about the same time, but were initially unrelated.  The first person to ask whether superstrings might be stretched to cosmic size and play the role of cosmic strings was Witten (1985).  His answer was in the negative, for several reasons.  First the string energy scale is close to the Planck energy, and so the superstring tension $\mu$ would be too high: $G\mu \gap 10^{-3}$, while it was clear even then that for cosmic strings $G\mu \lap 10^{-5}$, otherwise they would generate excessively large perturbations.  Secondly, any period of inflation would have diluted the numbers of pre-existing cosmic strings or superstrings to an unobservable level; only strings formed after or at the end of inflation could be observationally relevant.  Moreover, macroscopic Type-I strings would break up on a stringy time scale, and so would not form.  Macroscopic heterotic strings always appear as boundaries of axion domain walls, whose tension would force the strings to collapse rather than grow to cosmic  scales (Vilenkin \& Everett 1982). At the time of Witten's work no instability of long Type-II  strings was known, but it is now understood that Neveu-Schwarz 5-brane instantons (in combination with supersymmetry breaking to lift the zero modes) will produce an axion potential and so lead to domain walls (Becker \ea\ 1995). 

There was a dramatic change, however, after the `second string revolution' around 1995.  As we shall explain in the following sections, under the right conditions all these objections have disappeared.

\section{The tension issue}
\label{tension}
Two developments have made it possible to envisage strings with much lower tensions, compatible with observational limits on $G\mu$: large compact dimensions and warping.  These ideas are illustrated by two specific models.  (For a review see Myers \& Wyman 2009.) 

\subsection{Warping --- the \klmt model}
\label{klmt}
The idea of \emph{warping}, in which distances in the physical 4D spacetime are scaled by a warp factor depending on position in the internal space was first embodied in the model of Randall \& Sundrum (1999, 1999a).  Perhaps the model that has proven to be the best test bed for cosmic superstrings is that developed by Kachru \ea\ (2003a), and commonly referred to as the \klmt  model. Initially introduced as a way of realising inflation in string theory (Kachru \ea\ 2003) in a framework where all the moduli (the parameters defining the characteristics of the internal compact space) are stabilised (Giddings \ea\ 2002),  the formation of cosmic superstrings at the end of inflation provides an exciting additional feature. 

\begin{figure}
\begin{center}
\includegraphics[width=4in,angle=0]{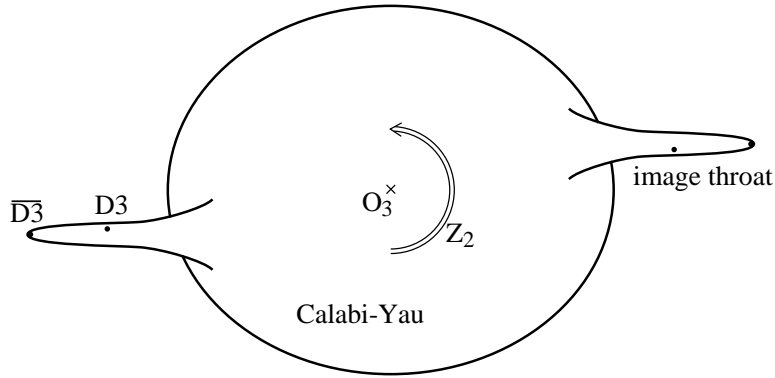}
\caption{Schematic picture of the \klmt geometry: a warped Calabi-Yau manifold with throats, identified under a ${\Z}_2$ orientifold (Copeland \ea\ 2004).}
\label{kklt-throat}
\end{center}
\end{figure}

The \klmt model is based on IIB string theory compactified on a Calabi-Yau manifold (illustrated schematically in Fig.~\ref{kklt-throat}). 
This is orientifolded by a ${\Z}_2$ symmetry (\ie, points related by this symmetry are identified, with singularities at its isolated fixed points). The spacetime
metric is warped, \ie, the metric of the physical four-dimensional spacetime is scaled by a factor depending on the position in the internal space:
\begin{equation}
ds^2 = e^{2A(x_\perp) }\eta_{\mu\nu} dx^\mu dx^\nu + ds_\perp^2\ ,
\end{equation}
where $x_\perp$ denotes the coordinates in the compact space.

Inflation is normally driven by a slowly varying scalar \emph{inflaton} field, whose potential energy is dominant during the process. In the \klmt model the inflaton field is the separation between a D3-brane and a $\OL{\mbox{D3}}$- (anti-D3)-brane, whose eventual annihilation when the separation vanishes leads to reheating.  The annihilation occurs in a region of large gravitational redshift, called the \emph{throat}, at the bottom of which $A$ attains its minimum value, ${\rm min}\{ e^{A(x_\perp)} \} = e^{A_0} \ll 1$, say, where $e^{A(x_\perp)}$ is normalized to be $O(1)$ in the bulk of the Calabi-Yau.
Note that in Fig.~\ref{kklt-throat} the covering manifold has two
throats which are identified under the ${\Z}_2$ rather than a single
throat identified with itself; thus there is no fixed point of the orientifold in the throat.

The redshift in the throat plays
a key role: both the inflationary scale and the scale of string tension, as
measured by a ten-dimensional inertial observer, are set by string
physics and are close to the four-dimensional Planck scale.
The corresponding energy scales as measured by a four-dimensional physicist
are then suppressed by a factor of $e^{A_0}$.

It is expected that the only objects that will be produced in any significant numbers will be
one-dimensional objects in the noncompact
dimensions lying entirely within the region of reheating (Jones \ea\ 2002, 2003, Sarangi \ea\ 2002).  The obvious candidates are then the F1-brane (fundamental IIB string) and
D1-brane, localized in the throat.  There is also another apparent possibility,
which however can be quickly disposed of.  In the \klmt model, the throat
has a Klebanov-Strassler (KS) geometry (Klebanov\ \& Strassler 2000), whose cross section  is
topologically $S^2 \times S^3$.  A D3-brane wrapped on the $S^2$ also
gives a string in four dimensions.  However, the $S^2$ collapses to a point at the bottom of the throat,  and so this string can break rapidly into small pieces.  The stability of the F- and D-strings will be discussed in Section~\ref{stability}.

Finally a caveat about the model we have been discussing. It should satisfy a number of
nontrivial constraints, including stability of the weak scale, stability of the moduli during inflation, and sufficient reheating.   The \klmt model appears to have a fatal problem with reheating related to the fact that there is a $U(1)$ gauge field on the stabilizing $\OL{\rm D3}$.  Being the only massless degree of freedom in the inflationary throat, it is the only one that couples directly to the inflationary fields.  Then almost all of the energy at reheating goes into these $U(1)$ gauge bosons rather than into the Standard-Model fields.  So this is not a fully satisfactory model. This isn't our main concern here; rather we have been attempting to show how strings can form. There are in fact more general constructions in which the problematic features of the \klmt model are absent. For example, the role of the $\OL{\rm D3}$-branes is to break supersymmetry, raising the supersymmetric anti-de Sitter vacuum to a state of approximately zero cosmological constant, but  there are other dynamical mechanisms that would accomplish the same thing. Similarly, there need not be an associated $U(1)$ field in general, thereby  eliminating the immediate problem with reheating.

\subsection{Large dimension models}
\label{large-dim}

In the early days of string theory, the only types of compactification that were seriously considered involved Calabi-Yau or orbifold compactifications (see Polchinski 1998 and Becker \ea\ 2007) where the size of the internal space was of the order $l\rms{s}$, but later the possibility emerged that the extra compact dimensions could be large (Arkani-Hamed \ea\ 1998,1999, Antoniadis \ea\ 1998).  Now in a Kaluza-Klein reduction, integrating the Einstein-Hilbert action $(1/16\pi G_{10})\int d^{10}x\, R$ over the six internal dimensions yields a relation between the 10- and 4-dimensional Newton constants $G_{10}$ and $G_4$, namely $G_{10} = V_6 G_4$, where $V_6$ is the volume of the six internal compact dimensions.  For Type-II strings, $16\pi G_{10} = (2\pi)^7 g\rms{s}^2 l\rms{s}^8$ (Polchinski 1998, Becker \ea\ 2007).  Thus the four-dimensional Planck length $l\rms{Pl}=\sqrt{G_4}$ is not fundamental, rather it is related to the string length $l\rms{s}$ by $l^2\rms{Pl} \sim g_s^2 (l\rms{s}^6 / V_6)l\rms{s}^2$.  Thus when $V_6 \gg l^6_s$ the four-dimensional Planck length is much smaller than the string length, whence the fundamental string tension is much smaller than $1/ l\rms{Pl}^2$: 
\beq
\label{4d-tension}
\mu_{\rm fun} \sim  \fr{1}{ l^2\rms{s}} \ll \fr{1}{ l\rms{Pl}^2} 
\sim \left(\fr{1}{ g\rms{s}^2}\fr{V_6 }{ l\rms{s}^6}\right)\fr{1}{ l\rms{s}^2}. 
\eeq

As yet no model involving large extra dimensions exists in which all moduli are stabilized, but it is still possible to investigate the stability of potential cosmic strings by first of all  fixing the moduli by hand. When this is done, a rich spectrum of strings is found depending on  the compactification scenario. In Sections \ref{inflation} and \ref{stability} below we will show how specific models of large extra dimensions can also lead to successful scenarios of inflation followed by the production of cosmic superstrings, with correspondingly small string tensions satisfying (\ref{tension-range})
(Jones \ea\  (2002,2003), Sarangi \& Tye (2002), Majumdar \& Davis 2002).  

\subsection{Strings from low-energy field theory}

There is always the possibility that our vacuum is well described by weakly coupled heterotic string theory and that braneworlds have played a minor role in reaching that position. If that were the case, we would expect that inflation and cosmic strings would have emerged out of the effective low-energy field theory (low, that is, in comparison to the Planck scale), with the possible strings forming including magnetic as well as electric flux tubes, the latter existing in strongly coupled confined theories (Witten 1985).  For example, Jeannerot \ea\ (2003) have shown that within the framework of Supersymmetric Grand Unified Theories (SUSY GUTs), cosmic strings are generically formed at the end of an inflationary era, providing a concrete realisation of how strings and inflation could co-exist in realistic particle physics inspired models. (For a nice review of cosmic strings arising from GUT theories see  Sakellariadou (2009)).

\section{The inflation issue}
\label{inflation}

A number of models have been proposed in which strings are formed at the end of a period of early-universe inflation. In D-brane inflation scenarios (Dvali \&Tye 1999, Burgess \ea\ 2001, Alexander 2002, Dvali \ea\ 2001) terminated by collisions between D3-branes and $\OL{\mbox{D3}}$-branes, D-strings are formed  (Burgess \ea\ 2001).   This is a special case of the production of strings during hybrid inflation (Yokoyama 1989, Kofman \ea\ 1996).  Jones \ea\  (2002,2003) and Sarangi \& Tye (2002) argued that these scenarios lead to the copious production of 
D-branes that are one-dimensional in the noncompact directions, and made the important observation that zero-dimensional defects (monopoles) and two-dimensional defects (domain walls) are not produced; either of these would have led to severe cosmological difficulties.

The reheating process during D3-$\OL{\mbox{D3}}$ annihilation can be described by a \emph{tachyon} field (one whose potential has a maximum at the origin, so that it starts to grow exponentially).
The D1-branes can be regarded as topological defects in this tachyon field (Sen 1999), and so these will be produced by the Kibble mechanism (Kibble 1976).  The F-strings do not have a classical description in these same variables, but in an $S$-dual
description they are topological defects and so must be produced in the
same way, implying that both D-strings and F-strings are expected to be produced at the end of brane inflation (Copeland \ea\  2004, Dvali \& Vilenkin 2003, 2004).  Of course, only one of the $S$-dual descriptions can be quantitatively valid, and if the string coupling is of order one then
neither is.  However, the argument depends only on causality, so it is plausible that it is valid for both kinds of string in all regimes.  All strings created at the end of inflation are at the bottom of the
inflationary throat, and they remain there because they are in a deep
potential well: their effective four-dimensional tension $\mu$ depends on the
warp factor at their location and their ten-dimensional tension $\bar\mu$ as
\begin{equation}
\mu = e^{2A(x_\perp) }\bar\mu \ .  \label{redten}
\end{equation}

Although only two types of string are produced, we end up with a much richer spectrum of string types (unlike conventional abelian cosmic strings), because when a D-string meets an F-string it can bind together rather than intercommute. In other words they can merge to form
a bound state known as a (1, 1)-string. Further binding leads to the formation of higher bound states, generally known as $(p, q)$-strings, which are composed of $p$ F-strings and $q$ D-strings where $p$ and $q$ are relatively prime integers. They were originally found (Schwarz 1995, Harvey \& Strominger 1995) using the $SL(2,{\Z})$ duality of the IIB superstring theory and are now interpreted  as bound
states of $p$ F1-branes and $q$ D1-branes (Polchinski 1995, Witten 1996).
Their tension in the ten-dimensional Type-IIB theory is (Schwarz 1995)
\begin{equation}
\label{pq-tension}
\bar \mu_{p,q} = \frac{1}{2\pi l^2_s} \sqrt{p^2 + \frac{q^2}{g_{\rm s}^2}}\ ,
\end{equation}
where $g_{\rm s}$ is the perturbative string coupling.

This result is valid for $(p,q)$-bound states in a flat ten-dimensional spacetime. It is modified when considering a cosmological background 
(Firouzjahi \ea\ 2006, Thomas\ \& Ward 2006).  For example in the case of  the collision of  cosmic superstrings in a warped throat, as would occur in models of warped brane inflation (Kachru \ea\ 2003a), the inflation takes place inside a  warped throat and cosmic superstrings produced at the end of inflation are located at the bottom of the throat. To be specific, consider $(p,q)$-strings in the KS throat which is a warped deformed conifold.  Here one finds for the tension of  a string at the tip of the throat (Firouzjahi \ea\ 2006)
\begin{equation}
\label{E}
\bar \mu_{p,q}=  \frac{e^{2A_0}}{2\pi l^2_s} \sqrt{ \frac{q^2}{g_s^2} + \frac{ M^2}{\pi^2}
 \sin^2 \left(  \frac{\pi p}{M}\right) },
\end{equation}
where $M$ is an integral quantum number (the Ramond-Ramond $F_{(3)}$ flux).
In the limit where $M\rightarrow \infty$, and the warping vanishes $A_0 \to 0$, this formula reduces to the tension of a $(p,q)$ string in a flat background given by Eq.~(\ref{pq-tension}).  

Strings are not an inevitable consequence of brane inflation. There are many models in which they do not appear, for example based on the condensation of an open string tachyon, or inflation based on closed string moduli. However, given the expectation that the  particle phenomenology associated with superstrings is expected to be rich, we may well expect there to be standard cosmic strings, with tensions well below the string scale emerging out of the supersymmetric grand unified theories (Jeanerrot \ea\ 2003). 

There are other proposals to form cosmic superstrings that do not rely on brane-inflation. The idea of a Hagedorn phase transition dates back to well before the second string revolution and has been used by a number of authors  to explain the distribution of strings in the early universe. The density of states of strings increases exponentially with their mass. At a particular scale (the Hagedorn scale, which is close to the string scale), there is a transition and all the available energy goes into creating long strings, as opposed to loops.  We might then expect some of these to survive until the present day (Polchinski 2004, Frey \ea\ 2006). If inflation is not required, then it is possible to start with a large universe in the Hagedorn phase (Nayari \ea\ 2006).    

Although it is clear that the formation of cosmic superstrings depends on the precise way in which inflation is occurring, what is important here is the fact that they can form at all and not be diluted away by the exponential expansion of the universe. So, given that they can form with a low tension and survive a period of inflation, can we make them stable enough to be cosmologically interesting? We now turn our attention to that question. 

\section{The stability issue}
\label{stability}

In his pioneering paper, Witten pointed out two types of instability the fundamental strings would have (Witten 1985). Of course, given that he also pointed out the inconsistency of the string tension with that required by the observation of the large-scale structure, it seemed just as well that they were unstable. The first type of instability is the fragmentation in open string theories (which we now think of in terms of the breakage of strings on space-filling 9-branes) and the second is confinement by axion domain walls with the resulting wall tension causing the string loops to collapse.  There are more {\em potential} instabilities we also need to be aware of, namely an effect similar to baryon decay and tachyon condensation. These involve technical calculations and we will not go into great detail here; the interested reader is recommended to read Myers\ \& Wyman (2009), Polchinski (2004,2005), or Copeland \ea\ (2004). 

Type-I strings were the first to be considered for their stability and it was shown that there is a constant rate per unit length for the string to break up forming a pair of new endpoints (Dai\ \& Polchinski 1989, Mitchell \ea\ 1989), the effect being that a macroscopic string would break up into microscopic strings which correspond to the lightest particle excitations.  The same process can occur for F-strings (fundamental Type-II strings) which can also end on D$p$-branes, leading to breakage. 
For complete instability this must be a space-filling D9-brane. If on the other hand we have $p<9$, then the D$p$-brane does not fill all the compact dimensions, leading to a suppression in the probability of  splitting as there can be a transverse separation between the strings and the D-branes, allowing metastable strings to exist in some models.  Although this argument applies primarily to Type I and II fundamental strings, using S and T duality arguments it can be extended also to other strings breaking on different branes (Copeland \ea\ 2004).  In general, for stability of macroscopic strings, there must be no space-filling branes, but this is a potential problem --- space-filling branes play an important role in model building; the low energy phenomenology seems to rely on their presence (Blumenhagen \ea\ 2005). This implies that macroscopic strings can have at best indirect interactions with the Standard-Model particles, gravity being the only mutual force. Otherwise they will break on space-filling branes. 

A second model of instability is that the strings become confined by axion domain walls. It is an instability first noted by Vilenkin\ \& Everett (1982) in the context of cosmic strings but applies equally well to the case of superstrings, particularly those formed by wrapping a $p$-brane on a compact $(p-1)$-cycle.  Axion fields then play an essential part in their construction.  Unless the tension of the wall is exceedingly small this will cause the string to rapidly disappear.

A third mode of instability involves  an analogue of `baryon decay'. Some of the string models have fluxes on their internal cycles, fluxes which lead to an interaction between the low-energy gauge fields and axions associated with the four-dimensional strings.  These fluxes can produce a new mechanism for string breakage. For example consider a type IIB compactification with a three-cycle ${\cal K}_3$ containing a nonvanishing Ramond-Ramond three-form flux
\begin{equation}
\label{bgd-flux}
\int_{{\cal K}_3} F_{(3)} = M.
\end{equation}
It is possible for a D3-brane to wrap on the same internal cycle. Being a localised particle in four dimensions, it is loosely referred to as a baryon; $M$ F-strings end on the baryon. For the case $M=1$, the fundamental strings can break by the production of baryon-antibaryon pairs (Copeland \ea\ 2004). However, for $M \geq 2$, this decay mechanism stops working, leading to stable strings. 

The final mode of decay is through the condensation of tachyons, responsible for the decay of an unstable D-brane. As it decays, the D-brane breaks up into ordinary quanta along its length, similarly to the case of the breaking on space-filling branes. The unstable D-brane is considered as being constructed from a D--$\OL{\mbox{D}}$-brane pair (Sen 1999), and the tachyon mediating the decay of the brane is an open string stretching between them. It is possible to suppress this decay by separating the brane-antibrane pair in the internal space, although even if that is done so that none of the open string modes are tachyonic, there remains the possibility of decay through nonperturbative tunneling processes.  

Overall, there are clear modes of decay for F and D strings of all types, and for them to remain viable candidates for cosmology conditions have to be in place to suppress the natural decay routes.  However, there are now models in which these conditions can be met, in particular the \klmt model, subject to conditions on the parameters, and some with large extra dimensions.
This is a complicated issue, which we cannot treat in detail here, but we can give a flavour of the physics responsible. Take for example the case of warped compactifications with throats. We have already established that one principal decay mode is through the breaking of strings on branes, implying that in general they should not co-exist in the same throat. Given that, and recalling that the strings and branes both feel a potential due to a gravitational redshift (warp factor) in the compact directions, to break the strings, they must tunnel from the inflationary throat to one of the other wells where they can end on a brane.  Depending on the model being considered, the rate for this this can be very slow indeed implying the existence of effectively stable strings (Copeland \ea\ 2004).  In particular for the strings produced in inflationary models driven by D3--$\OL{\mbox{D3}}$-brane collisions, such as the \klmt model, there are three possible outcomes (Copeland \ea\ 2004): (a) no strings will exist if all decay routes are in operation; (b) D1-branes only (or fundamental strings only) persist if only some instabilities are present; (c) $(p,q)$-strings with an upper bound on p exist, if the instabilities are under control. 

For the case of wrapped branes in a Type-II compactification, Jones \ea\  (2002,2003) and Sarangi \& Tye (2002) found a rich family of stable strings, with string tensions in the range 
\begin{equation}
\label{tension-range}
10^{-12} \leq G\mu \leq 10^{-6},
\end{equation}
with the preferred value around $10^{-7}$.  In obtaining this bound the authors worked on the assumption that the CMB anisotropies are due to quantum fluctuations of the inflaton field as opposed to any other mechanism. This then allows the string tension to be related to the observed quantities arising in brane inflation.  The details of the allowed strings are quite technical and the reader is referred to the original papers for the full description. As a brief summary, the authors consider brane inflationary scenarios where the string scale is close to the GUT scale implying that the post-inflation brane-world models are supersymmetric at the GUT scale. In the 10-dimensional superstring theory, the cosmic strings living in 4-dimensional spacetime are D$p$-branes with $p-1$ dimensions wrapped on a compact cycle and one spatial dimension along one of the 3 large spatial dimensions.  Considering a  typical Type-IIB orientifold model compactified on  the $T^6/{\Z_2}$ orientifold, where
the ${\Z_2}$ reflects $k$ coordinates, the authors obtain the number of 
possible stable (after inflation) configurations of branes of different
dimensionality in 10 dimensions, compactified on the six manifold. There are many of them including long lived D1-branes, F-strings and hence $(p,q)$-strings. Although the Standard-Model branes present may allow some of these strings to break,  those that are wrapped on different cycles from the 
Standard-Model branes, for example, will remain stable.
 
\section{Intercommuting properties}
\label{reconnect}
In this section we begin to address perhaps the most important question concerning cosmic superstrings. How can we differentiate a network of them from the more traditional field-theory-based cosmic strings? We have seen in earlier sections that it is possible to form cosmic superstrings from string theory and that they can be long-lived (cosmologically), survive a period of inflation and have a macroscopic length. They bring with them two particular features that may help us distinguish the two types. They are a reduced probability of intercommuting (Jackson \ea\ 2005) and the formation of junctions in the $(p,q)$-string networks (Sen 1998). We will concentrate on these two features in this section.

\subsection{Reconnection probabilities} 
\label{prob}
We have already seen that field-theory simulations of cosmic strings indicate that the probability of intercommuting is essentially unity, $P=1$, with only ultra-relativistic strings being able to pass through each other without reconnecting, a result true of both global (Shellard 1987) and local (Matzner 1989) cosmic strings. The case for superstrings is different however, as pointed out in great detail by Jackson \ea\ (2005). Earlier, Polchinski (1988) had shown that in the context of 
string perturbation theory, for fundamental strings, the reconnection probability depends primarily on the string coupling constant $g_s$ and is of order $g^2_s$, allowing it of course to be much less than one. 

Jackson \ea\ (2005) extended this approach, analysing the collisions  between all possible pairs of  fundamental and Dirichlet strings, as well as  their $(p, q)$ bound states. The strength of the interaction between the colliding strings  depends on the details of the compactification, the relative velocity of the strings, their intersection angle and crucially on $g_s$. Averaging over angles and velocities they obtained estimates for the intercommuting probability for three scenarios A,B and C. Case A involves the \klmt model in which quantum fluctuations over the $S^3$ of the Klebanov-Strassler throat solution have been averaged over. Case B is where the fluctuations  in the throat do not fill out the $S^3$ and case C is the situation for models of large extra dimensions.  The reconnection probability $P$ depends strongly on the type of strings and on the details of the compactification, with full details given by Jackson \ea\ (2005). The analysis shows that the D-D string reconnection probability is in the range $0.1 \lesssim P \lesssim 1$, whereas for F-strings $10^{-3} \lesssim P \lesssim 1$. As far as $(p,q)$ strings are concerned, the probability can be larger or smaller than for single strings.  For the case where there are stable cosmic strings for more than one $(p,q)$ value, the reconnection probabilities of different types of strings are very important in determining what happens to the network. The fact that the probability of reconnection $P$ is generally below unity means that networks of F- or D-strings can in principle be distinguished from gauge-theory strings. If $P\sim 10^{-3}$, that would have a large effect on the behavior of string networks. 

An important consequence of a reduction in intercommutation rates is that  the density of long strings has to increase because loop formation becomes less efficient as a mechanism for energy loss. Quite how the density changes with probability is a matter of some debate. Semi-analytical arguments based on scaling suggest that the density of long strings should scale as $\rho_{\rm str} \propto P^{-1}$ (Sakellariadou 2005). However, numerical studies of string networks where the intercommutation probability is reduced have shown that the strings tend to cross microscopically many times during an intercommutation event, the result being that for $P\geq 0.1$ there is very little enhancement of $\rho_{\rm str}$ whereas for $P\leq 0.1$, one finds $\rho_{\rm str} \propto P^{-\gamma}$ with $\gamma \sim 0.6$ (Avgoustidis\ \& Shellard 2006).

\subsection{Bound states} 
\label{bound-states}

When strings of two different types cross they cannot intercommute in the
same way as usual.  Rather they can produce a pair of trilinear vertices connected
by a segment of string (Figure~\ref{zip}). For example, the crossing of a $(p,q)$ string and a $(p',q')$ string can
produce a $(p+p',q+q')$ string or a $(p-p',q-q')$ string. Only for $(p',q') = \pm (p,q)$ is the usual intercommutation possible. We saw earlier that the D3/$\OL{\rm D3}$-brane inflation model can lead to the formation of a network of $(p,q)$-strings. For the case of the ten-dimensional Type-IIB theory the tension is given by Eqn.~(\ref{pq-tension}) although the specific form depends on the underlying nature of the string compactification (see for example Eqn.~(\ref{E})). In general it depends on the square root of a function of $p$ and $q$.
\begin{figure}
\begin{center}
\includegraphics[width=4in,angle=0]{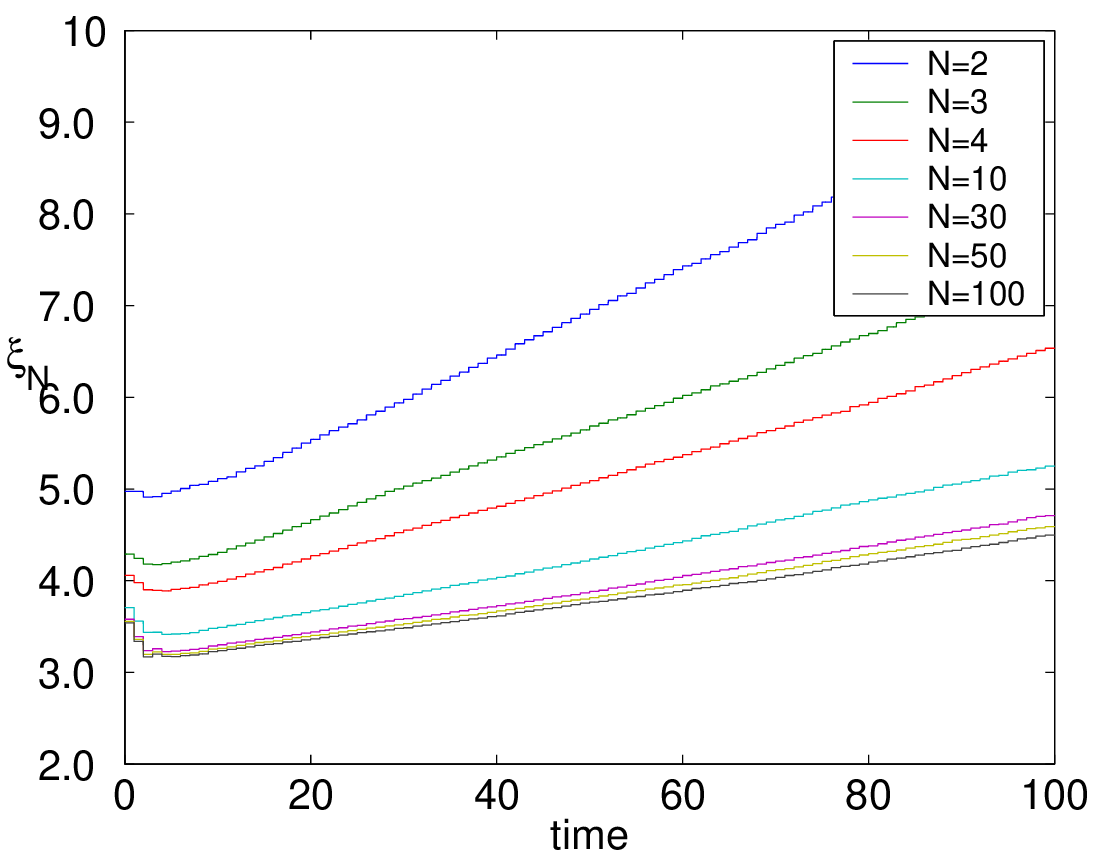}
\caption{The evolution of the typical string length-scale $\xi_N$ as a function of N. Scaling corresponds to $\xi_N \propto t$ (Copeland\ \& Saffin 2005).}
\label{string-scaling}
\end{center}
\end{figure}

Over the past few years an alternative approach has emerged to determine the properties of a network of $(p,q)$ strings, which should be a good approximation at least for the lowest lying tension states. That has been the development of field-theory analogues, either combinations of interacting abelian models (Saffin 2005, Rajantie \ea\ 2007) or through non-abelian models (Spergel\ \& Pen 1997, McGraw 1998, Copeland\ \& Saffin 2005). In both cases the networks that form admit trilinear vertices, hence junctions where the usual intercommutation properties of the strings no longer apply.

It has been suggested that if $P$ is large (\ie\ $P \sim 1$), the strings can in principle freeze into a three-dimensional network different from the usual scaling solution. It is likely however that complicated networks will form initially but that a scaling regime is eventually reached, in which the ratios of lengths of different kinds of strings are constant. Evidence for this has been provided both analytically (Tye \ea\ 2005, Avgoustidis\ \& Shellard 2006) and by numerically modelling the network with field-theory networks (Copeland\ \& Saffin 2005, Hindmarsh\ \& Saffin 2006).

The conclusion is that a network of bound strings will reach a scaling solution with the lowest lying states, the F-, D- and (1,1)-strings, having higher number densities compared to the higher-tension strings. This is because the unbinding of higher-tension strings is favoured kinematically compared with the binding processes.  As an example of the scaling solutions which are achieved from complicated initial networks of strings see Figure \ref{string-scaling} which is taken from the field-theory simulations of Copeland\ \& Saffin (2005). Considering a field theory containing $N$ different species of string, they study the effect of non-intercommuting events due to two different species crossing each other, and look for evidence of scaling indicated by the characteristic length scale evolving so that it eventually scales linearly with the proper time $t$, i.e. $\xi_N \propto t$. 

A key aspect of string theory is the existence of string dualities which relate different string models, hence can relate different types of string that are formed. The $(p,q)$-strings form a mulitiplet under a discrete SL(2,$\Z$) symmetry of the ten-dimensional Type-IIB theory (Schwarz 1995). This means that an F-string (a (1,0)-string) can be mapped to a general $(p,q)$-string by an appropriate application of an SL(2,$\Z$) transformation, leading to the formation of many types of string, not just the basic F-string. Of course this does not mean they would all form; that would depend on the detailed dynamics of the particular model being considered, but in principle they could form leading to a rich structure of low-tension bound cosmic superstrings. 

The kinematics of strings that can form junctions is a fascinating area of research which has only recently started receiving attention (Copeland \ea\ 2006,2007, Davis \ea\ 2008).  A number of important features have emerged through studies of the collisions of Nambu-Goto strings with junctions at which three strings meet.  One is that the exchange to form junctions cannot occur if the strings meet with very large relative velocity. For the case of non-abelian strings, rather than passing through one another they become stuck in an X configuration (Copeland \ea\ 2007), in each case the constraint depending on the angle at which the strings meet, on their relative velocity, and on the ratios of the string tensions. Calculations of the average speed at which a junction moves along each of the three strings from which it is formed yield results consistent with our earlier observation that junction dynamics may be such as to preferentially remove the heavy strings from the network leaving a network of predominantly
light strings. 

Copeland \ea\ (2008) extended this discussion to include the formation of three-string junctions between $(p,q)$-cosmic superstrings, which required modifications of the Nambu-Goto equations to take account of the additional requirements of flux conservation. Investigating the collisions between such strings they showed that kinematic constraints analogous to those found previously for collisions of Nambu-Goto strings apply here too. Extending their analysis to the \klmt-motivated model of the formation of junctions for strings in a warped space, specifically with a KS throat, they showed that similar constraints still apply with changes to the parameters taking account of the warping and the background flux. 

A number of these  constraints that have emerged from analysing the modified Nambu-Goto equations have been checked through numerical simulations of field-theory strings which can also lead to junction formation (Sakellariadou\ \& Stoica 2008, Salmi \ea\ 2008, Bevis\ \& Saffin 2008, Bevis \ea\ 2009). In most cases good agreement has been obtained with the analytical predictions, although the results of Urrestilla \& Vilenkin (2008) show some surprising features, and of course some important differences emerge due to the fact that field-theory strings have interactions between them that are not found in the Nambu-Goto case, hence some of the detailed numbers differ. For example, Salmi \ea\ (2008) found, when investigating the collision of strings numerically in the framework of themabelian Higgs model in the Type-I regime, that strings can effectively pass through each other when they meet at speeds slightly above the analytically obtained critical velocity permitting bound-state formation. This is due to reconnection effects that are beyond the scope of the Nambu-Goto approximation.

\subsection{Problems of stable relics}
Analytical modelling (Tye \ea\ 2005, Avgoustidis\ \& Shellard 2008) suggests that if too many species form then the overall number density of strings would increase dramatically above the usual scenarios.  The knock-on effect would be that the average velocity of strings in the network would go to zero, leading to the formation of a cosmologically forbidden frustrated network of strings --- forbidden because such a network would quickly come to dominate the energy density of the universe and act as dark energy with effective equation of state $w\equiv p/\rh=-\frac13$ which is ruled out by recent observations (Spergel \ea\ 2007).

As mentioned earlier, the appearance of monopoles or domain walls would be observationally unacceptable.  But even without these, another possible problem was pointed out by Avgoustidis\ \& Shellard (2005). It involves the possibility that, in winding around a compact extra dimension in a way that forbids the loop to vanish, the strings themselves can form new stable remnants they termed {\it cycloops}.  If they do form in a network the strings responsible for them would have to have incredibly small tensions, $G\mu \lesssim 10^{-18}$, so they would be impossible to detect with current or planned experiments.

There also remains the possibility of stable loops of superconducting string forming {\it vortons} (Davis\ \& Shellard 1989) which are stabilised by their angular momentum and cannot be radiated away classically. If formed from a first-order transition these would be disastrous cosmologically for all $G\mu > 10^{-20}$ (Martins\ \& Shellard 1998). However these conclusions rely on the superconducting currents being the same as those of the gauge groups of the Standard Model, which is not very likely as we have already established that for cosmic superstrings to be stable against breakage on space-filling branes seems to require that the cosmic superstrings have only indirect, \eg\ gravitational, interactions with the Standard-Model particles. Of course it is possible to turn these problems on their heads and tune the parameters so that the remnant loops actually play the role of dark matter. 

\section{Observational signatures of cosmic strings and cosmic superstrings}
\label{observation}

The most exciting aspect of cosmic superstrings must be the fact that they could leave cosmological signatures for us --- they may be out there waiting to be detected in the large-scale features of our universe. We now turn our attention to this, but first a word of warning. Most of the predictions to date are based on approaches developed for the original cosmic strings, which did not possess the richness of the cosmic superstrings. Therefore they should be treated with a degree of caution; the full calculations remain to be performed. However, there are features which may allow us to distinguish between the two and we will highlight these in this section where appropriate. 

\subsection{Gravitational effects of strings}
\label{graveffects}

Both cosmic strings and cosmic superstrings experience gravitational interactions, which offer  the best possibility for observationally confirming their existence.  Depending on the precise nature of the strings, other types may interact in different ways, and offer distinct possibilities for detection. We will briefly comment on this later in this section.  One of the strings' gravitational effects would be to generate density perturbations in the universe, and at one time it was thought that this might provide an explanation for the origin of the primordial perturbations from which galaxies and stars have developed.  However observations of the cosmic microwave background (CMB) by COBE and WMAP have shown conclusively that this effect can at most account for a small fraction of the total power, up to 10\% (Pogosian \ea\ 2009, Bevis \ea\ 2008).  The theory of inflation provides a very good fit to the observed power spectrum, in particular the acoustic peaks, whereas cosmic strings alone would yield a featureless spectrum with a single broad peak.  This imposes an upper limit on the string tension: $G\mu \lesssim 2.1 \X 10^{-7}$ (Fraisse 2007).

The gravitational field around a cosmic string (or cosmic superstring) is quite unusual.  Around a straight, local cosmic string, spacetime is actually locally flat; the gravitational acceleration towards the string is zero (Vilenkin 1981).  This is because the tension and energy per unit length are equal, but tension acts as a negative source of the gravitational field.  However, spacetime is not globally flat, but essentially cone-shaped, as though a wedge with apex along the string had been removed and the two faces stuck together. 

The deficit angle $\de$ is related to the dimensionless constant $G\mu$ introduced in Section \ref{evolution}; in fact $\de=8\pi G\mu$.  Explicitly, if the string is along the $z$ axis, the spacetime metric around it is
 \beq ds^2=dt^2-dz^2-d\rh^2-(1-8G\mu)\rh^2d\vp^2, \eeq
which is flat in terms of the modified azimuthal coordinate $\vp'=(1-4G\mu)\vp$, ranging from 0 to $2\pi-8\pi G\mu$.  This constant $G\mu$ is the key parameter determining whether or not there are observable effects of strings.  From (\ref{mu}) we see that for strings created at an energy scale $T\rms{c}$, $G\mu$ is of order $(T\rms{c}/M\rms{Pl})^2$, where $M\rms{Pl}$ is the Planck mass ($=G^{-1/2}\ap 10^{19}$~GeV).  In particular for strings created at a GUT transition, we expect $G\mu$ to be of order $10^{-6}$ or $10^{-7}$.  The deficit angle is given explicitly by 
 \beq 
 \de=5.2\left(\fr{G\mu}{10^{-6}}\right)\text{~arcsec}. 
 \eeq

Note that spacetime is only locally flat around \emph{straight} strings.  If they are wiggly, with many kinks, then the effective energy per unit length $U$ will be larger than $\mu$ and the effective tension $T$ will be smaller; typically $TU=\mu^2$ (Carter 1990, Vilenkin 1990).  In that case, in addition to the deficit angle, there is a net gravitational acceleration towards the string.
\begin{figure}
\begin{center}
  \includegraphics[width=2.5in,angle=0]{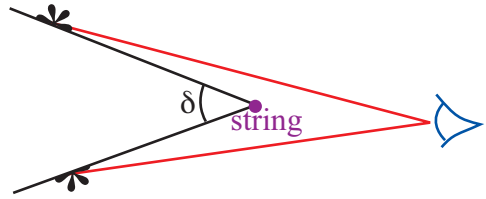}
\caption{Lensing of a distant source by a string.}
\label{lens}
\end{center}
\end{figure}

\subsection{Lensing}
\label{lensing}

The most obvious effect of this geometry is gravitational lensing.  Light from a distant source may reach us round either side of a cosmic string (see Fig.~\ref{lens}), thus creating a double image with a separation angle of order $\de$.  If the string is straight, then since each ray passes only through locally flat spacetime, the images are not magnified, in sharp contrast to gravitational lensing by other kinds of objects, though this is not precisely true if the strings are wiggly.  Thus a very characteristic signature of cosmic-string lensing is the appearance of a pair of images of very similar magnitude, and without the distortions seen in most cases of gravitational lensing.  Moreover one would expect to see other such pairs in the neighbourhood, ranged along the path of the string.  There have in fact been two reported observations that seemed at first sight to be possible examples of this phenomenon (Cowie \& Hu 1987, Sazhin \ea\ 2003), though in both cases further study has shown that they are probably merely pairs of surprisingly similar objects that happen to lie close together (Hu 1990, Sazhin \ea\ 2006).  Given that the typical distance $\xi$ between strings is a significant fraction of the horizon distance, and that imaging would only occur for objects lying beyond the string and within a very narrow band of width at most of order an arc-second or so, to have found such a pair might seem rather fortuitous.  Morganson \ea\ (2009) recently presented limits based on a survey of possible lensed optical images.

Gravitational lensing also offers a way of detecting cosmic superstrings, in particular networks of $(p,q)$-strings. As we have seen in Section~\ref{reconnect} these strings can form junctions at which three strings meet, opening up the possibility of new lensing signatures of a three-way lensing junction (Shlaer\ \& Wyman 2005, Brandenberger \ea\ 2008) as light passes by them.  In principle knowing the angles in the pattern would allow for the relative tensions of the strings to be determined, and this could be tested against the predictions of a set of $(p,q)$-strings where the three tensions are linked. 

There has also been discussion of the possibility of detecting very small, rapidly oscillating loops by their microlensing effects (Dyda \& Brandenberger 2007, Chernoff \& Tye 2007), although others have concluded that this is unlikely to succeed (Kuijken \ea\ 2007).  Future radio interferometers may be able to detect lensing by loops, thereby offering an efficient way of observing or placing limits on them (Mack \ea\ 2007).  A characteristic feature of the perturbations induced by cosmic strings is that they include vector as well as scalar and tensor modes.  One consequence of this would be rotation of the images of distant galaxies, which could be detected in weak lensing surveys (Thomas \ea\ 2009).

\subsection{Effect on the cosmic microwave background}
\label{CMB}

Another very characteristic signature of cosmic strings would be their effect on the CMB (Kaiser \& Stebbins 1984).  If a string is moving transversely across the sky with velocity $v$, there will be a red-shift of radiation passing ahead of it, and a blue-shift of radiation passing behind it.  Thus one should see a line discontinuity in the CMB temperature, of order $\de T/T\sim G\mu v$.  Since the strings are moving fast, this temperature difference should in principle be detectable.  Unfortunately the temperature variation occurs only over a very narrow angular band (of order $\de$ wide) and existing detectors such as WMAP do not yet have the required resolution (Jeong \& Smoot 2007).  However it is possible that it could be seen at the South Pole Telescope, provided that $G\mu > 5.8 \X 10^{-8}$ (Stewart \& Brandenberger 2009).  It is also conceivable that the Planck instrument might do so. 

The vector-mode perturbations induced by cosmic strings will in turn lead to polarization of the CMB, and this may turn out to be one of the best ways of detecting cosmic strings, in particular via the B-mode polarization (Pogosian \& Wyman 2008%, Urrestilla \ea\ 2008
) which the Planck satellite will be capable of detecting.

The case of small-scale anisotropies generated by strings with junctions has been considered by Brandenberger \ea\ (2008), who conclude that a distinctive signature in this case is line discontinuities joined at a point  for an observer looking at the surface
of last scattering, with the details of the image depending on the orientations of the strings and
the direction of  motion of the junction with respect to the line of sight. 

\subsection{Gravitational radiation}
\label{gravrad}

One of the most promising ways of detecting the presence of cosmic strings and cosmic superstrings is via the gravitational radiation they emit.  If, as is widely accepted, emission of gravitational radiation is the primary mechanism of energy loss from strings, then they will have contributed a substantial amount to the gravitational-wave background in the universe.  An important role in the emission of gravitational radiation is played by the appearance of cusps on the strings.  These are points where in the Nambu-Goto approximation the unit vectors $\bp$ and $\bq$ of (\ref{pq}) satisfy $\bp=\bq$.  At such points, the string momentarily reaches the speed of light.  Consequently, the gravitational radiation emitted in the neighbourhood of this event is very strongly beamed in the direction of the string velocity. Damour \& Vilenkin (2005) have shown that the cusps yield a very characteristic gravitational-wave signature which, depending on the intercommuting probability $P$ and number of cusps per loop oscillation, could be detected for string tensions as low as $G\mu \sim 10^{-10}$, placing them well within range of the predicted values for cosmic superstrings arising out of brane inflation (see Eqn.~\ref{tension-range}). In particular as $P$ decreases, the density of strings in a string network increases leading to the production of more gravitational waves and a higher strain signal. Existing detectors are not yet sensitive enough to have a good chance of seeing this signal, though LIGO has been used to exclude some regions of parameter space (Abbott \ea\ 2009).  Whether it will be visible to LIGO2 or LISA will again depend strongly on the value of $G\mu$ and on various other parameters (Polchinski 2007). 

There are strong limits on the gravitational wave background stemming from observations of millisecond pulsars, which serve as extremely accurate clocks.  Any gravitational radiation present will disturb the timing of the pulses.  So the observed regularity of the pulses over more than two decades places a firm upper limit on the energy density in gravitational waves (Hobbs \ea\ 2009).  This then translates into an upper limit on the value of the cosmic string tension:
 \beq G\mu < 1.5\X 10^{-8}c^{-3/2}, \eeq
where $c$ is the mean number of cusps on each loop.  This is probably the most stringent current limit on $G\mu$.  The actual value of the limit depends strongly on two other parameters: the ratio $\al=l/t$ of the typical loop size to the time and the intercommuting probability $P$ described above (which we recall is close to 1 for most ordinary cosmic strings, but can be substantially less for cosmic superstrings). Recently a number of authors have turned their attention to the specific cases of gravitational radiation from strings with junctions which are more applicable to the cosmic superstring case  (Siemens \ea\ 2006, 2007, Brandenberger \ea\ 2009a, Binetruy \ea\ 2009, Jackson\ \& Siemens 2009), but much needs to be done to really establish their impact on the gravitational wave signature. 

Gravitational radiation emitted by cosmic strings will also induce B-mode polarization in the CMB, but this will be swamped by the much larger effect of the vector modes mentioned above.

\subsection{Other observations}
\label{otherobs}

There are several other possible ways in which cosmic strings might reveal their presence.  High-energy particles may be emitted by strings from the regions around their cusps, conceivably providing an explanation for the observed very-high-energy cosmic rays (Brandenberger \ea\ 2009).  Cusps might also perhaps induce shock waves that could generate gamma-ray bursts (Berezinsky 2001).  Although not generic, it is possible that cosmic superstrings could be superconducting (Witten 1985a), carrying massless degrees of freedom charged under an unbroken gauge symmetry (Polchinski 2004).  Certain kinds of strings could yield a mechanism for generating primeval galactic magnetic fields (Davis \& Dimopoulos 2005, Gwyn \ea\ 2008).  Yet another effect might be the generation of early reionization.  Olum \& Vilenkin (2006) estimate that to avoid conflict with observation this would require $G\mu < 3 \X 10^{-8}$.

\subsection{Observational features particular to cosmic superstrings}
\label{unique-features}

Imagine the exciting situation where some signature of cosmic strings is seen. Could we determine whether what has been observed are cosmic superstrings as opposed to say cosmic strings arising in GUT theories? There are two principal ways we may be able to do so. 

The first is somewhat indirect and makes use of the differences in the respective reconnection probability $P$ discussed in Section~\ref{reconnect}(\ref{prob}) whilst the second is based on the fact that cosmic superstrings can form junctions and bound states as discussed in Section~\ref{reconnect}(\ref{bound-states}). This in turn leads to new features in the string network as compared to the `vanilla' type of string which is normally discussed. The fact that the reconnection probabilities for F-F, D-D and F-D strings can be substantially less than unity implies that the amount of string in the network must be increased by some factor related to the probability. As discussed earlier, a consequence of increasing the density of strings is that there can be a dramatic increase in the gravitational-wave signal produced by the network (Damour\ \& Vilenkin 2005), in the most favourable case making it possible that LIGO could well soon start seeing many cusps. Of course, increasing the density of string in the universe has the knock on effect of making the existing bounds on the string tension even stronger, because there would be more string to see, hence the allowed string tension must go down. If the probability is dramatically different from unity, the standard field-theory value, then given enough observations of the gravitational-wave signal, it should be possible to determine both the string tension $\mu$ and $P$, hence to rule out standard field-theory cosmic strings as the source. However, if the probability differs only slightly from unity, then it will be much more difficult to distinguish the two sources, although it is the case that cosmic strings arising out of GUT models tend to have $P=1$ as a robust result, implying that a result differing from unity would be significant. 

The  formation of networks  of bound state $(p,q)$-strings discussed in  Section~\ref{reconnect}(\ref{bound-states}) could be a smoking gun for cosmic superstrings. Both analytic and numerical modeling of comparable networks suggest that they will reach a scaling solution with an enhanced string density, and, given that, it could well be possible to eventually have enough information to distinguish F-D string networks from other types. In the context of brane inflation, these $(p,q)$ networks can be considered as a fairly generic outcome because the inflation usually takes place in a warped throat region, and it is in such a situation that we find both F and D-strings forming, hence $(p,q)$ strings with a tension given by Eqn.~\ref{E} for example. Unfortunately networks with junctions are not unique to bound states of F and D strings. Mulitple types of string can arise in field theories leading to complicated networks with junctions, although the particular spectrum of Eqn.~\ref{E} could well be unique. Even if that is the case, duality relations exist between the various string solutions meaning there will be gauge theory strings that are very hard to differentiate from these F- and D- strings.  Looking at the network as a whole, if it was possible to measure the tensions of many strings in the network, it would be possible see whether the distribution of tensions fitted the form predicted by a $(p,q)$ network, in particular was it of the form $\mu \sim \sqrt{f(p^2,\,q^2)}$? Realistically, such precision is unlikely, and given that we expect only the lowest values of $p$ and $q$ to be excited, it could well be that a tuned field theory will mimic what is found. However, there is little doubt that either of these two results (three-way lensing and multiple tensions) would generate a lot of excitement in the string community and prompt more detailed investigation.  

There are other possibilities that could provide indirect evidence of cosmic superstrings (for details see Myers\ \& Wyman 2009). They include the formation of `baryons' or beads on a string network, and strings ending on monopoles which can be considered as beads with only a single string attached, and which can produce observable gravity wave bursts (Leblond \ea\ 2009).  There has also been interest recently in the possibility that the equivalent of semilocal strings (Vachaspati\ \& Ach\'ucarro 1991) could be found in D3/D7 models of brane inflation (Urrestilla \ea\ 2004).

\section{Conclusions}
\label{conc}
Many early-universe scenarios predict the appearance of cosmic strings or superstrings.  The most promising perhaps is brane-world inflation.  Models in which inflation is driven by a collision between  D3- and $\OL{\mbox{D3}}$-branes generically predict the formation of strings of various types, F- and D-strings and composite $(p.q)$-strings.  The details of which strings are stable or metastable are complicated but the conditions are not overly restrictive. Even in models where inflation does not involve branes, cosmic superstrings may be formed by other mechanisms.  Moreover, ordinary cosmic strings may appear later in the evolutionary process.

There are many possible ways by which cosmic strings or superstrings may be detected.  Their direct lensing effect may be observable if $G\mu$ is reasonably close to the present observational upper bound.  Loops offer the possibility of being seen if their tension is low enough, because they last longer, hence their number density will increase. This is particularly the case if they were to cluster near our galaxy (Chernoff\ \& Tye 2007). In fact these authors claim that the planned astronomical survey mission GAIA could observe strings with tensions as low as $G\mu \geq 10^{-10}$ because the loops generate microlensing events that will be within GAIA's reach. Small-scale observations of the CMB using the Atacama Cosmology Telescope (ACT) and the South Pole Telescope (SPT) may reveal the characteristic discontinuities created by cosmic strings for tensions satisfying $G\mu \geq 10^{-7}$ (Fraisse \ea\ 2008). The fact that strings can directly source vector mode perturbations at last scattering implies that even light strings can generate strong B-mode polarisation in the CMB, providing a signal which is distinguishable from the type of spectrum expected from inflation (Pogosian \& Wyman 2008, Bevis \ea\ 2008). Current experiments searching for  characteristic patterns in the B-mode polarisation, include BICEP, PolarBear, and Spider and these have just been joined by Planck in the search. Gravitational radiation emitted from cusps, as well as from kinks or junctions, may be detectable by future experiments, such as LISA and remain probably the most sensitive test for cosmic strings and cosmic superstrings, as they could detect loops with tensions as low as $G\mu \geq 10^{-13}$ (Damour\ \& Vilenkin 2005).  Meanwhile, the introduction of Pulsar Timing Arrays will provide ever more constraining limits on the allowed string tensions as will direct gravity wave searches like LIGO2 or LISA.

Success in any of these observations would trigger huge interest, and allow us to plan further observations to probe the detailed nature of the strings, in particular to distinguish ordinary cosmic strings from superstrings.  A positive result would yield a lot of information about the nature of the underlying fundamental theory, information that would be very hard to gather in any other way.

\end{document}